\documentclass[aps,reprint,nofootinbib]{revtex4-1}
\pdfoutput=1

\usepackage{setspace}

\usepackage{booktabs}

\usepackage{graphicx}
\usepackage{amsmath}
\usepackage{amsfonts}
\usepackage{amssymb}
\usepackage{rotating}

\usepackage{tikz}

\usepackage{float}
\usepackage{multirow}
\usepackage{slashed}
\usepackage{hyperref} 
\hypersetup{colorlinks=true, linkcolor=red, urlcolor=red, citecolor=red}
\usepackage[normalem]{ulem}
\usepackage[utf8]{inputenc}

\newcommand{\AddrAHEP}{%
  AHEP Group, Institut de F\'{i}sica Corpuscular --
  C.S.I.C./Universitat de Val\`{e}ncia, Parc Cient\'ific de Paterna.\\
 C/ Catedr\'atico Jos\'e Beltr\'an, 2 E-46980 Paterna (Valencia) - Spain}
 
\newcommand{\AddrMPI}{Max-Planck-Institut f\"ur Kernphysik,\\ Saupfercheckweg 1, 69117 Heidelberg, Germany}

\begin{document}

\title{\color{red}{Two-Higgs-Doublet Models with a Flavored $\mathbb{Z}_2$}}

\author{S.\ Centelles Chuli\'a}
\email{salcen@ific.uv.es}
\affiliation{\AddrAHEP}

\author{W.\ Rodejohann}
\email{rodejoha@mpi-hd.mpg.de}
\affiliation{\AddrMPI}

\author{U.~J.\ Salda\~na-Salazar}
\email{saldana@mpi-hd.mpg.de}
\affiliation{\AddrMPI}

\begin{abstract}
\noindent
Two Higgs-doublet models usually consider an ad-hoc $\mathbb{Z}_2$ discrete
symmetry to avoid flavor changing neutral currents. 
We consider a new class of two Higgs-doublet models where $\mathbb{Z}_2$ 
is enlarged to the symmetry group ${\cal{F}}\rtimes \mathbb{Z}_2$, i.e. an inner semi-direct product of a discrete symmetry group  ${\cal{F}}$ and $\mathbb{Z}_2$. 
In such a scenario the symmetry constrains the Yukawa interactions but goes unnoticed by the scalar sector. In the most minimal scenario, $\mathbb{Z}_3 \rtimes \mathbb{Z}_2 = D_3$,  
flavor changing neutral currents mediated by scalars are absent at tree and one-loop level, 
while at the same time predictions to quark and lepton mixing are 
obtained, namely a trivial CKM matrix and a PMNS matrix (upon introduction of three heavy 
right-handed neutrinos) containing maximal atmospheric mixing.  
Small extensions allow to fully reproduce mixing parameters, including cobimaximal mixing in the lepton sector 
(maximal atmospheric mixing and a maximal $CP$ phase).  
\end{abstract}

\maketitle

\section{Introduction}
\label{sec:intro}
\noindent
The discovery of a Higgs boson with a mass of $m_h \simeq 125$ GeV has opened the door
to the possibility of having in Nature multiple fundamental scalars. In principle, nothing
forbids their proliferation. Nonetheless, the amount of parameters dramatically increases,
both in the Yukawa and scalar sector. Here we consider a simple extension to the 
standard model (SM) by only introducing a second Higgs doublet (2HDM) with quantum numbers identical to the SM Higgs, and three right-handed neutrinos to generate active neutrino masses. 
Furthermore, we mainly focus on the problem of fermion mixing by first 
adopting the common 2HDM framework
with natural flavor conservation (NFC)~\cite{Paschos:1976ay,Glashow:1976nt}, achieved through a $\mathbb{Z}_2$ reflection symmetry. Then, we add \textit{flavor} 
to it via the enlargement of the symmetry group in a very particular manner, 
${\cal F} \rtimes \mathbb{Z}_2 $. This denotes an inner semi-direct product of a 
discrete symmetry group ${\cal{F}}$ and a $\mathbb{Z}_2$ symmetry. 
The non-Abelian nature of the enlarged symmetry group then strongly reduces the number of Yukawa couplings, thus providing a more predictive theory. Moreover, the ad-hoc nature of the $\mathbb{Z}_2 $ is explained as a part of a larger group\footnote{There are other possibilities to explain the ad-hoc $\mathbb{Z}_2 $, for instance by linking it to the remnant symmetry of a spontaneously broken $U(1)$, see e.g.\ \cite{Ko:2012hd,Campos:2017dgc,Camargo:2018klg}}.

To understand the need for the $\mathbb{Z}_2$ symmetry we briefly sketch its impact. 
In a general setup, one may immediately write the Yukawa Lagrangian for a given fermion
\begin{align}
    -{\cal L}_Y \supset \overline{\psi}_L \left( {\bf Y}^\psi_1 \Phi_1 + {\bf Y}^\psi_2 \Phi_2 \right) \psi_R + \text{ H.c.}, 
\end{align}
where $\psi_R$ or $\psi_L$ are three-dimensional vectors in flavor space
 each denoting a weak singlet or a weak doublet, respectively,
and $\psi$ representing any of the four fermion types, 
$\psi= q_u, q_d, \ell, \nu$.
Notice that the Higgs doublets must be replaced by their charge-conjugate fields, $\widetilde{\Phi}_k = i \sigma_2 \Phi^*_k$,
for the up-type quark and neutrino cases\footnote{For brevity, we left aside the Majorana option. 
However, we return to it later.}.
If the neutral components of both scalar doublets acquire a vacuum expectation value (VEV), 
$\langle \Phi_1^0 \rangle = v_1$ and $\langle \Phi_2^0 \rangle = v_2$, both Yukawa matrices
contribute to the fermion masses and mixing. It is clear that diagonalization of the 
mass matrix,  
\begin{equation}
    {\bf M} = v_1 {\bf Y}_1 + v_2 {\bf Y}_2 \; ,
\end{equation}
cannot mean, in general, diagonalization of
the individual Yukawa matrices. This brings about dangerous tree-level flavor-changing-neutral-currents (FCNC).
To avoid them it is sufficient to assume NFC by introducing a $\mathbb{Z}_2$ symmetry and
by assigning a single scalar doublet for a given fermion species such that
only one of the two Yukawa matrices contributes to the mass matrix. This is, 
the scalar fields transform under the discrete symmetry such that
\begin{align}
    \Phi_1 \rightarrow -\Phi_1 \; , \quad
    \Phi_2 \rightarrow +\Phi_2 \; ,
\end{align}
while the left-handed quarks and leptons transform trivially and the right-handed parts transform appropriately. The different assignment possibilities lead to 
four non-equivalent types of 2HDMs\footnote{The type X and Y are also called the lepton-specific and flipped
scenarios, respectively.}:
\begin{itemize}
    \item Type I: All charged fermions couple to $\Phi_2$.
    \item Type II: $q_d$ and $\ell$ couple to $\Phi_1$ and $q_u$ to $\Phi_2$.
    \item Type X: $q_u$ and $q_d$ couple to $\Phi_2$ and $\ell$ to $\Phi_1$.
    \item Type Y: $q_u$ and $\ell$ couple to $\Phi_2$ and $q_d$ to $\Phi_1$.
\end{itemize}
Other possibilities are the Type III which is the general 2HDM with all couplings permitted and 
the inert doublet model where $\Phi_2$ couples to all fermions while $\Phi_1$
has no VEV thus leaving unbroken the $\mathbb{Z}_2$ symmetry and providing a viable dark 
matter candidate. Although other approaches, such as  Yukawa-alignment~\cite{Pich:2009sp} or 
singular-alignment~\cite{Rodejohann:2019izm}, may also avoid tree-level FCNC, here we only focus
on those 2HDM employing the discrete symmetry $\mathbb{Z}_2$. Attempts to add flavor to 2HDMs have already been made, e.g.~\cite{Alves:2017xmk,Alves:2018kjr,Altmannshofer:2018bch}. However, our approach offers an alternative and novel way to consider the NFC theories as the starting point to build minimal extensions where the patterns in fermion mixing are taken as a guide for new physics. Note that
the non-equivalence nature between the four types comes from the fact that
each framework ends up having different effective Yukawa couplings of the
fermions to the various scalar particles; for a thorough discussion on various phenomenological and
theoretical aspects of 2HDMs see Ref.~\cite{Branco:2011iw}.

On the other hand, a general feature shared by the four different types (I, II, X, and Y) is
the $\mathbb{Z}_2$-invariant scalar potential given by
\begin{align} \label{eq:2HDMZ2}
\begin{split}
V_\text{2HDM}^{\mathbb{Z}_2}  = & \sum_{x=1,2} \left[ m^2_{xx} (\Phi^\dagger_x \Phi_x)
+\frac{\lambda_x}{2} (\Phi^\dagger_x \Phi_x)^2 \right] \\
& + \lambda_3 (\Phi^\dagger_1 \Phi_1) (\Phi^\dagger_2 \Phi_2) 
+\lambda_4 (\Phi^\dagger_1 \Phi_2) (\Phi^\dagger_2 \Phi_1)  \\
&+\frac{1}{2} \left[ \lambda_5(\Phi^\dagger_1 \Phi_2)^2 + \lambda_5^*(\Phi^\dagger_2 \Phi_1)^2 \right]  .
\end{split}
\end{align}
The hermiticity condition of the potential leaves $\lambda_5$ as the only complex coefficient 
while the rest, $m_{11}^2, m_{22}^2$, and $\lambda_{1,2,3,4}$, are real. There are in total 
eight real parameters. However, not all of them are physical. A phase redefinition can make $\lambda_5$
real and only seven parameters are physical. Note that our potential has explicitly become $CP$-symmetric. 

No matter the amount of Higgs doublets one employs, the full mass matrix for any given fermion 
is parametrized by nine complex parameters. The initial arbitrariness may then be reduced via weak-basis
transformations (unitary transformations leaving invariant the kinetic terms), but not enough to
claim predictivity. In the mass basis, for either quarks or leptons, one has six fermion masses and four (six for Majorana neutrinos) 
mixing parameters, plus arbitrary Yukawa couplings. The flavor sector thus gives to the SM and its
extensions (without symmetries) the highest amount of arbitrariness. It is only when symmetries are
introduced that the initial arbitrariness can be drastically reduced. Here we intend to explore
the effect of symmetries in the flavor sector such that we find correlations among the quark
and lepton mixing parameters.   \\

The paper is organized as follows: in Sec.~\ref{sec:flavor}, we discuss
the meaning of adding flavor to $\mathbb{Z}_2$. Next, in Sec.~\ref{sec:minimal},
we provide the most minimal scenario realizing the features of our approach.
Also we highlight the main differences when compared to the four types 
of 2HDMs. Thereafter, in Sec.~\ref{sec:AddRealScalar}, we take the incompleteness of fermion mixing in our  simple model as a hint to the presence of additional
new physics and introduce a flavor doublet of real scalar gauge singlets. Finally, in Sec.~\ref{sec:concl}, we conclude. Some technical details are delegated to appendices.

\section{Adding flavor to $\mathbb{Z}_2$}
\label{sec:flavor}\noindent
We are interested in those finite symmetry groups, ${\cal G}$, which can be written as an inner
semi-direct product of an arbitrary group ${\cal F}$ and $\mathbb{Z}_2$,
\begin{align}
        {\cal G} = {\cal F} \rtimes \mathbb{Z}_2 \;.
\end{align}
There are in fact many examples of such groups (see Ref.~\cite{Ishimori:2010au} for more details): 
$D_N  = \Delta (2N) \simeq \mathbb{Z}_N \rtimes \mathbb{Z}_2$,
$\Sigma (2N^2) \simeq (\mathbb{Z}_N \times \mathbb{Z}^\prime_N) \rtimes \mathbb{Z}_2$, 
$\Sigma(24) \simeq \mathbb{Z}_2 \times \mathbb{Z}_6 \rtimes \mathbb{Z}_2$, etc.

The main property of this kind of groups is that they contain two 
one-dimensional irreducible representations (denoted singlets),
which behave exactly as if we only had a $\mathbb{Z}_2$ symmetry. Thus, by assigning
each Higgs doublet to one of these singlets, we are mimicking in the scalar
sector any of the NFC models with a $\mathbb{Z}_2$ symmetry. On the other hand, the 
non-Abelian nature of the symmetry only impacts the Yukawa interactions, 
thus providing a way to approach the problem of mixing while simultaneously
tackling minimal scalar extensions to the SM.

An additional feature of this approach is the following. Since the number of Higgs doublets in a 
theory restricts the maximum group order of allowed symmetries ('realizable symmetries') 
that would otherwise imply massless Goldstone bosons~\cite{Ivanov:2011ae},
then by implementing symmetry groups as here proposed we avoid these constrictions. 

Let us take as a first example the Klein group given by $\mathbb{Z}_2 \rtimes \mathbb{Z}_2$.
It is the smallest possibility within this approach. 
It has four elements and four irreducible representations (irreps):
${\bf 1}_{++}$, ${\bf 1}_{+-}$, ${\bf 1}_{-+}$, and ${\bf 1}_{--}$.
However, as it is still an Abelian group its effect on the Yukawa
couplings is only of reduction but not of relation. For example,
we could assign the Higgs doublets as $\Phi_1 \sim {\bf 1}_{--}$ and 
$\Phi_2 \sim {\bf 1}_{++}$
while the third, second, and first fermion families  as
${\bf 1}_{-+}$, ${\bf 1}_{+-}$, and ${\bf 1}_{++}$, respectively. 
In return the mass matrix for Dirac fermions would take the
generic form
\begin{align}
    {\bf M} = 
    \begin{pmatrix}
        y_1 v_{++} & y_4 v_{--} & 0 \\
        y_5 v_{--} & y_2 v_{++} & 0 \\
        0 & 0 & y_3 v_{++}
    \end{pmatrix}  ,
\end{align}
where $\langle \Phi_{1}^0 \rangle = v_{--}$ and $\langle \Phi_{2}^0 \rangle = v_{++}$. Therefore, although we have reduced the number of complex parameters from nine to five,
we yet have no predictions except for the fact that we only expect mixing 
between the first two generations. 
Nevertheless, it demonstrates that the combination of the flavor-safe $\mathbb{Z}_2$ with an additional group will simplify the Yukawa sector. Going to the non-Abelian case will result in predictive scenarios, and we will study a very minimal approach in what follows. 

\section{The minimal case: $\mathbb{Z}_3 \rtimes \mathbb{Z}_2$}
\label{sec:minimal}\noindent
The smallest non-Abelian finite group has six elements and is denoted by $D_3 \equiv \mathbb{Z}_3 \rtimes \mathbb{Z}_2$. 
This dihedral group describes the symmetrical properties of an equilateral triangle\footnote{$D_3$ is isomorphic 
to $S_3$, the group describing the permutations of three indistinguishable objects.}. It has three irreducible representations: two singlets ${\bf 1}_+, {\bf 1}_-$, and one doublet ${\bf 2}$. 
The product rules can be found in Appendix~\ref{app:Rules}.

Although different assignments between the $D_3$ irreps 
and the fermion fields could be done, here we opt to consider
\begin{align} 
\begin{split}
	Q_{L,3} \sim {\bf 1}_+ \;, \quad Q_{L,D} =
	\begin{pmatrix}
	Q_{L,1} \\
	Q_{L,2}
	\end{pmatrix} \sim {\bf 2} \;, \\ 
	u_{R,3} \sim {\bf 1}_+ 
	\;, \quad 
	u_{R,D} = \begin{pmatrix}
	u_{R,1} \\
	u_{R,2}
	\end{pmatrix} \sim {\bf 2} \;, \\ 
	d_{R,3} \sim {\bf 1}_- 
	\;, \quad 
	d_{R,D} = \begin{pmatrix}
	d_{R,1} \\
	d_{R,2}
	\end{pmatrix} \sim {\bf 2} \;,
\end{split}
\end{align}
whereas in the lepton sector, 
\begin{align} 
\begin{split}
	E_{L,1} \sim {\bf 1}_+ \;, \quad E_{L,D} =
	\begin{pmatrix}
	E_{L,2} \\
	E_{L,3}
	\end{pmatrix} \sim {\bf 2} \;, \\ 
	e_{R,1} \sim {\bf 1}_- 
	\;, \quad 
	e_{R,D} = \begin{pmatrix}
	e_{R,2} \\
	e_{R,3}
	\end{pmatrix} \sim {\bf 2} \;, \\ 
	N_{R,1} \sim {\bf 1}_- 
	\;, \quad 
	N_{R,D} = \begin{pmatrix}
	N_{R,2} \\
	N_{R,3}
	\end{pmatrix} \sim {\bf 2} \;.
\end{split}
\end{align}
We are motivated to this choice, as we will see, because the dominant contributions to quark and lepton mixing are the Cabibbo and atmospheric angle, correspondingly.

Recall that the scalar sector should be assigned to 
\begin{align}
	\Phi_1 \sim {\bf 1}_- \qquad \text{and} \qquad \Phi_2 \sim {\bf 1}_+ \;.
\end{align} 
The neutral component of both Higgs doublets acquires a VEV, spontaneously
breaking the electroweak symmetry; we denote them as
\begin{equation}
   v_1 \equiv \langle \Phi_1^0 \rangle  \; \qquad \text{and} \qquad
   v_2 \equiv \langle \Phi_2^0 \rangle \;.
\end{equation}
Note we are using the convention $v^2 = v_1^2 + v_2^2 = (174 \text{ GeV})^2$.

The $\mathbb{Z}_3 \rtimes \mathbb{Z}_2$-symmetric Yukawa Lagrangian is
\begin{equation}
    -{\cal L}_Y = {\cal L}_Y^Q + {\cal L}_Y^E
\end{equation}
with
\begin{align}
\begin{split}
    {\cal L}_Y^Q = & \; y_t \overline{Q}_{L,3} \widetilde{\Phi}_{2} u_{R,3} 
    + y_b \overline{Q}_{L,3} {\Phi}_{1} d_{R,3} \\ 
    & +
    y_1^u [\overline{Q}_{L,D} u_{R,D}]_{-} \widetilde{\Phi}_1 +
    y_2^u [\overline{Q}_{L,D} u_{R,D}]_{+} \widetilde{\Phi}_2 \\
    & + 
    y_1^d [\overline{Q}_{L,D} d_{R,D}]_{-} {\Phi}_1 +
    y_2^d [\overline{Q}_{L,D} d_{R,D}]_{+} {\Phi}_2 \\
    & + \text{ H.c.},
\end{split}
\end{align}
and
\begin{align}
\begin{split}
    {\cal L}_Y^E = & \; y_{\nu 1} \overline{E}_{L,1} \widetilde{\Phi}_{1} N_{R,1} 
    + y_e \overline{E}_{L,1} {\Phi}_{1} e_{R,1} \\ 
    & +
    y_1^\nu [\overline{E}_{L,D} N_{R,D}]_{-} \widetilde{\Phi}_1 +
    y_2^\nu [\overline{E}_{L,D} N_{R,D}]_{+} \widetilde{\Phi}_2 \\
    & + 
    y_1^e [\overline{E}_{L,D} e_{R,D}]_{-} {\Phi}_1 +
    y_2^e [\overline{E}_{L,D} e_{R,D}]_{+} {\Phi}_2 \\
    & + \frac{1}{2} M_1 \overline{N^c_{R,1}} N_{R,1} +
    \frac{1}{2} M_2 [\overline{N^c_{R,D}} N_{R,D}]_+
    \\ & + \text{ H.c.},
\end{split}
\end{align}
where $[\quad]_k = \{ {\bf 1}_+, {\bf 1}_-, {\bf 2} \}$
represents one of the three possible outputs from the ${D}_3$ tensorial product.
Also notice that we are now assuming Majorana neutrinos by virtue of a
standard seesaw.

In the quark sector, the resulting Yukawa matrices take the form
\begin{align}
\begin{split} \label{eq:massmatquark}
    {\bf \Delta}_1 = & \, \begin{pmatrix}
        0 & y^u_1 & 0 \\
        -y^u_1 & 0 & 0 \\
        0 & 0 & 0
    \end{pmatrix} , \quad
    {\bf \Delta}_2 =
    \begin{pmatrix}
        y^u_2  \, e^{i \gamma_u} & 0 & 0 \\
        0 & y^u_2  \, e^{i \gamma_u} & 0 \\
        0 & 0 & y_{t}
    \end{pmatrix} , \\
    {\bf \Gamma}_1 = & \, \begin{pmatrix}
        0 & y^d_1 & 0 \\
        -y^d_1 & 0 & 0 \\
        0 & 0 & y_b
    \end{pmatrix} , \quad
    {\bf \Gamma}_2 =
    \begin{pmatrix}
        y^d_2  \, e^{i \gamma_d} & 0 & 0 \\
        0 & y^d_2   \, e^{i \gamma_d}& 0 \\
        0 & 0 & 0
    \end{pmatrix} ,
\end{split}    
\end{align}    
while in the lepton sector we have
\begin{align}
\begin{split} \label{eq:massmatlep}
    {\bf \Pi}_1 = & \,\begin{pmatrix}
        y_e & 0 & 0 \\
        0 & 0 & y_1^e  \\
        0 & -y_1^e & 0
    \end{pmatrix} , \quad
    {\bf \Pi}_2 =
    \begin{pmatrix}
        0 & 0 & 0 \\
        0 & y^e_2 \, e^{i \gamma_e} & 0 \\
        0 & 0 & y^e_2 \, e^{i \gamma_e}
    \end{pmatrix} , \\
    {\bf \Omega}_1 = & \, \begin{pmatrix}
        y_{\nu 1} & 0 & 0 \\
        0 & 0 & y_1^\nu \\
        0 & -y_1^\nu & 0
    \end{pmatrix} ,\quad
    {\bf \Omega}_2 = 
    \begin{pmatrix}
        0 & 0 & 0 \\
        0 & y^\nu_2  \, e^{i \gamma_\nu} & 0 \\
        0 & 0 & y^\nu_2  \, e^{i \gamma_\nu}
    \end{pmatrix} ,
\end{split}    
\end{align}    
where all the parameters are real and positive and where we have taken $\{ y_1^{u},y_1^d,y_{t},y_b,y_1^e,y_e \} \in \Re^+$ without loss of generality. 
All Dirac mass matrices satisfy 
\begin{equation} \label{eq:Diracmasses}
    {\bf M} = v_1 {\bf \Xi}_1 + v_2 {\bf \Xi}_2 \; ,
\end{equation}
where ${\bf \Xi} = {\bf \Gamma}, {\bf \Delta}, {\bf \Pi},$ and ${\bf \Omega}$.
Each mass matrix has three complex parameters  and
possesses the feature of being diagonalisable by the same transformation
that brings to diagonal form its individual Yukawa matrices. It is this
property that guarantee the absence of FCNC at tree level and it represents
an explicit realization of the singular alignment ansatz~\cite{Rodejohann:2019izm}.

Note how we end up, in the quark sector, with only eight real
parameters, six of which correspond to the six quark masses while the other two,
being complex phases, are forced to be nearly $\pm \pi/2$ due to
the phenomenological observation of hierarchical fermion masses. We return 
to this point later.

The effective Majorana neutrino mass matrix can be computed from the standard seesaw formula, 
${\cal M}_\nu =  - {\bf M}_\nu {\cal M}_R^{-1} {\bf M}_\nu^T$, and is found to be diagonal:
\begin{align} \label{eq:MajMasses}
	{\cal M}_\nu =  -\begin{pmatrix}
		\frac{(y_{\nu 1} v_1)^2}{M_1} & 0 & 0 \\
		0 & \frac{(y_2^\nu)^2 v_2^2 - (y_1^\nu)^2 v_1^2}{M_2} & 0 \\
		0 & 0 & \frac{(y_2^\nu)^2 v_2^2 - (y_1^\nu)^2 v_1^2}{M_2}
	\end{pmatrix} .
\end{align}
Here  ${\cal M}_R = \text{diag}(M_1,M_2,M_2)$, which is a consequence of 
the $D_3$ flavor symmetry.
The mass matrix has a mass degeneracy between the two 
neutrino states, $\nu_{L,2}$ and $\nu_{L,3}$, while, since it is diagonal, 
it does not contribute to the mixing. 

Towards studying the phenomenology of this scenario we note that complex matrices of the form 
\begin{align} {\bf m} = a {\mathbb{I}} +
    \begin{pmatrix}
        0 & b \\
        -b & 0
    \end{pmatrix}  ,
\end{align}
are brought to diagonal shape via a maximal bi-unitary transformation
\begin{align} \label{eq:complxphases}
 {\bf u}_L = \frac{1}{\sqrt{2}} 
    \begin{pmatrix}
       1 & \pm i\\
       \pm i & 1 
    \end{pmatrix} , \qquad
 {\bf u}_R = \begin{pmatrix}
       e^{i\gamma_{1}} &  0\\
       0 &   e^{i\gamma_{2}}
    \end{pmatrix} \cdot {\bf u}_L \,,
    \end{align}
    that is, 
    \begin{align}
    {\bf u}_L \cdot {\bf m} \cdot {\bf u}_R^\dagger = \begin{pmatrix}
       |a \mp i b|  & 0\\
       0 & |a\pm ib| 
    \end{pmatrix}  ,
\end{align}
with $\gamma_1 = \text{arg}(a \mp i b)$ and $\gamma_2 = \text{arg}(a \pm i b)$ implying real and positive masses. The choice of the signs will depend on the ordering of the masses. The singular values of such a matrix $\bf{m}$ are given by
\begin{align}
    m_{1,2} = |a \pm i b | = \sqrt{ |a|^2 + |b|^2 \mp 2 |a||b| \sin \rho } \;,  
\end{align}
where $\rho = \text{arg} (a)-\text{arg} (b)$. Moreover, note that if the parameters $a$ and $b$ are taken to be real ($\rho = 0$) then the masses would be degenerate. In particular, if $\rho$ is in the first quadrant then $\rho \in [\text{ArcSin}\left(\frac{m_2^2-m_1^2}{m_2^2+m_1^2}\right), \, \pi/2]$. The transformations $\rho \rightarrow \pm\rho + \pi$ and $\rho \rightarrow -\rho$ will lead to the same masses as $\rho$. Additionally, when the masses are hierarchical, $m_2 \gg m_1$, the allowed
interval for $\rho$ shrinks to $\rho \in [\pi/2 - 2m_1/m_2+{\cal O}(m_1^3/m_2^3), \, \pi/2]$,
essentially implying that $\rho \simeq \pm \pi/2$. 
We have chosen the off-diagonal Yukawas to be real and positive without loss of generality. 
Therefore the complex phase of the diagonal Yukawas is found to be $\gamma_f \simeq \pm\pi/2$. 

With these results in mind and looking at the form of the mass matrices of the charged fermions shown in Eqs.~\eqref{eq:massmatquark}-\eqref{eq:Diracmasses} we can extract the masses and mixing parameters:
\begin{equation}
\begin{gathered} \label{eq:masses}
    m_t  = y_t v_2 \;, \quad m_b = y_b v_1 \;, \quad m_e = y_e v_1  \;, \\
    m_{c,u}   =  |y^u_2 v_2 \pm  \, y^u_1 v_1| \;, \quad 
    m_{s,d} =  |y^d_2 v_2 \pm  \, y^d_1 v_1| \;, \\
    m_{\tau,\mu} =  |y^e_2 v_2 \pm  \, y^e_1 v_1| \;,
\end{gathered}
\end{equation}
with the Majorana neutrino masses as given in Eq.~\eqref{eq:MajMasses}.
The quark Yukawa couplings can now be generically fixed to (defining $\tan \beta = v_2/v_1$) 
\begin{align} \label{eq:hierarchy}
    y_1^{u(d)} = \frac{m_{c(s)} - m_{u(d)}}{2 v \cos \beta} \quad \text{and} \quad
    y_2^{u(d)} = \frac{m_{c(s)} + m_{u(d)}}{2 v \sin \beta} \;. 
\end{align}
An alternative solution exists when one exchanges $y_1^f \leftrightarrow y_2^f$. Similarly for the charged leptons, 
\begin{align} \label{eq:lepyuks}
    y_1^{e} = \frac{m_{\tau} - m_{\mu}}{2 v \cos \beta} \quad \text{and} \quad
    y_2^{e} = \frac{m_{\tau} + m_{\mu}}{2 v \sin \beta} \;,
\end{align}
and again it is possible to exchange $y_1^e \leftrightarrow y_2^e$.

Turning to fermion mixing, we can parametrize the relevant diagonalization matrices in terms of the complex rotation  matrices ${\bf U}_{ij}(\theta, \phi)$, which are defined as
\begin{equation}
    {\bf U}_{12}(\theta, \phi) = \begin{pmatrix}
        \cos \theta & \sin \theta e^{-i \phi} & 0 \\
        -\sin \theta e^{i \phi} & \cos \theta &0 \\
        0 & 0 & 1 \\
    \end{pmatrix} ,
\end{equation}
 and similarly for $U_{13}$ and $U_{23}$. Then, the mixing matrices for the up and down quarks and for the charged leptons are simply given by
 \begin{align}
     {\bf U}_L^u = {\bf U}_{12}(\pi/4, \pm \pi/2) \; , \\
     {\bf U}_L^d = {\bf U}_{12}(\pi/4, \pm \pi/2)  \; ,\\
     \label{eq:UL} {\bf U}_L^e = {\bf U}_{23}(\pi/4, \pm \pi/2) \; .
 \end{align}
 We obtain for the Cabibbo-Kobayashi-Maskawa (CKM) and Pontecorvo-Maki-Nakagawa-Sakata (PMNS) matrices
\begin{equation}
\begin{gathered}
    {\bf V}^{(0)}_\text{CKM}  =  {\bf U}^u_L ({\bf U}^d_L)^\dagger = 
    \begin{pmatrix}
        1 & 0 & 0\\
        0 & 1 & 0\\
        0 & 0 & 1
    \end{pmatrix} , \\
    \label{eq:PMNSini} {\bf U}^{(0)}_\text{PMNS}   = 
    {\bf U}^e_L ({\bf U}^\nu_L))^\dagger =
    \begin{pmatrix}
        1 & 0 & 0\\
        0 & \frac{1}{\sqrt{2}} & \pm \frac{i}{\sqrt{2}}\\
        0 & \pm \frac{i}{\sqrt{2}} & \frac{1}{{\sqrt{2}}}
    \end{pmatrix}  ,
\end{gathered}
\end{equation}
where one of the signs of the PMNS matrix is realized when Eq.~\eqref{eq:lepyuks} applies and the opposite when $y_1^e \leftrightarrow y_2^e$.

This is, by enlarging $\mathbb{Z}_2$ to $\mathbb{Z}_3 \rtimes \mathbb{Z}_2$, we are now able 
to predict  trivial mixing in the quark sector and a maximal atmospheric mixing angle in the lepton sector. 
There is also a maximal $CP$ violation phase, which is unphysical if the angles $\theta_{12}$ and $\theta_{13}$ remain $0$, but it will become important later. These features have to be understood as the dominant characteristics of this model at leading order.
Its incompleteness points to further investigation on how the model should be
extended, see Sec.\ \ref{sec:AddRealScalar}.

\subsection{FCNC}
\label{ssec:FCNC}\noindent
There are no tree level FCNC since all Yukawa matrices are simultaneously diagonalisable. 
However, at the one-loop level quantum corrections could induce 
misalignment in the different Yukawa matrices and generate FCNC. 
To check this effect we employ the formulas obtained for a theory with $N-$Higgs doublets~\cite{Ferreira:2010xe}
and given in Appendix~\ref{app:RGE}. It is straightforward to see that for our particular model in all cases the one-loop renormalization-group-equations
may only give place to flavor-conserving terms\footnote{As we are only interested in finding
flavor-violating structures, we have not considered the quantum corrections to the VEVs.}
\begin{equation}
    16\pi^2 \mu \frac{d}{d\mu}{\bf \Xi}_a \propto {\bf \Xi}_a \;,
    \qquad (a=1,2),
\end{equation}
where $\mu$ is the renormalization scale and
${\bf \Xi} = {\bf \Gamma}, {\bf \Delta}, {\bf \Pi},$ and ${\bf \Omega}$. More details can be found in Appendix~\ref{app:RGE}\footnote{Due to the fact that we are employing the standard seesaw, FCNC with heavy sterile neutrinos are sufficiently suppressed
and are  not discussed here.}.

\subsection{Nonuniversal charged fermion-scalar couplings}
\label{ssec:couplings}\noindent
In order to find the couplings between the charged fermions and the Higgs
scalars we need to move both of them to their mass basis. In our case, only the latter are
still in the symmetry adapted basis. We first introduce their notation
\begin{align}
    \Phi_1 = \begin{pmatrix}
        \Phi_1^+ \\
        v_1 + \frac{\phi_1^0+i\varphi_1^0}{\sqrt{2}}
    \end{pmatrix} , \quad
    \Phi_2 = \begin{pmatrix}
        \Phi_2^+ \\
        v_2 + \frac{\phi_2^0+i\varphi_2^0}{\sqrt{2}}
    \end{pmatrix}  .
\end{align}
Since the scalar potential is $CP$-symmetric there are states with definite $CP$-odd and $CP$-even quantum numbers. This allows one to write two independent mass
matrices
\begin{align}
    {\bf M}^2_{CP\text{-even}} = \begin{pmatrix}
        2 v_2^2 \lambda_2 & 2 v_1 v_2 \lambda_{345} \\
        2 v_1 v_2 \lambda_{345} & 2v_1^2 \lambda_1
    \end{pmatrix},
\end{align}
where $\lambda_{345} = \lambda_3 + \lambda_4 + \lambda_5$, and
\begin{align}
    {\bf M}^2_{CP\text{-odd}} = \begin{pmatrix}
        2 v_2^2 \lambda_5 & -2 v_1 v_2 \lambda_{5} \\
        -2 v_1 v_2 \lambda_{5} & 2v_1^2 \lambda_5
    \end{pmatrix}.
\end{align}
The first case can be brought to diagonal form by means of the orthogonal transformation
\begin{align}
    \begin{pmatrix}
        h \\
        H
    \end{pmatrix} =
    \begin{pmatrix}
        c_\alpha & s_\alpha \\
        -s_\alpha & c_\alpha 
    \end{pmatrix}
    \begin{pmatrix}
        \phi_2^0 \\
        \phi_1^0
    \end{pmatrix},
\end{align}
with $\tan 2\alpha = 2 v_1 v_2 \lambda_{345} /(v_1^2 \lambda_1 - v_2^2\lambda_2)$, 
while the second one by 
\begin{align}
    \begin{pmatrix}
        G^0 \\
        A
    \end{pmatrix} =
    \begin{pmatrix}
        c_\beta & s_\beta \\
        -s_\beta & c_\beta 
    \end{pmatrix}
    \begin{pmatrix}
        \varphi_2^0 \\
        \varphi_1^0
    \end{pmatrix} .
\end{align}
Here the latter angle of rotation satisfies $\tan \beta = v_2 / v_1$ and $G^0$ is the neutral
pseudo-Goldstone boson to be `eaten' by the $Z$ mass. Similarly, one has for 
the charged scalars a mass matrix
\begin{align}
    {\bf M}_\text{charged}^2 =
    \begin{pmatrix}
     2v_2^2 (\lambda_4 + \lambda_5) & -2v_1 v_2 (\lambda_4 + \lambda_5) \\
     -2v_1 v_2 (\lambda_4 + \lambda_5) & 2v_1^2 (\lambda_4 + \lambda_5)
     \end{pmatrix} ,
\end{align}
diagonalised by the same rotation as for the $CP$-odd neutral scalars,
\begin{align}
    \begin{pmatrix}
        G^+ \\
        H^+
    \end{pmatrix} =
    \begin{pmatrix}
        c_\beta & s_\beta \\
        -s_\beta & c_\beta 
    \end{pmatrix}
    \begin{pmatrix}
        \Phi_2^+ \\
        \Phi_1^+
    \end{pmatrix} .
\end{align}

The Yukawa Lagrangian related to the interactions to the neutral scalars is 
\begin{align}\label{eq:couneu}
    -{\cal L}_Y \supset  \sum_{f} \frac{m_f}{(246 \text{ GeV})} \left( \xi_h^f \overline{f}f h 
    + \xi_H^f \overline{f}f H -i \xi_A^f \overline{f}\gamma_5 f A   \right) ,
\end{align}
while the one related to the interactions to the charged scalar is
\begin{align}\label{eq:coucha}
\begin{split}
    -{\cal L}_Y \supset & - H^+ \frac{[{\bf V}^{(0)}_\text{CKM}]_{ij}}{(246 \text{ GeV})} \overline{u}_i\left( m_i \xi^{H^+}_{q_u} {\bf P}_L+ m_j  \xi^{H^+}_{q_d} {\bf P}_R \right) d_j \\
    & - H^+ 
    \frac{m_\ell}{(246 \text{ GeV})}  \xi^{H^+}_{\ell} \overline{\nu}_{L,i} \ell_{R,j}  + \text{ H.c.},
\end{split}
\end{align}
with ${\bf P}_{L,R} = (1\mp\gamma_5)/2$.  In order to compare this expression
to that appearing in conventional 2HDMs we have assumed, for the moment, massless neutrinos.

We find that an important distinction between this framework with 
typical 2HDMs with NFC (see Table~\ref{tab:couplings1}) is that 
fermion couplings become nonuniversal, see Table~\ref{tab:couplings2}.
Furthermore, those fermions which initially talk to both Higgs doublets ($\Phi_{1,2}$)
have the following couplings 
\begin{align} \label{eq:newcouplings}
\begin{split}
   f_{\pm}(\alpha,\beta,y_1,y_2) = & \frac{y_2 c_\alpha \mp y_1 s_\alpha}{y_2 s_\beta \pm y_1 c_\beta} \; , \\
   g_{\pm}(\alpha,\beta,y_1,y_2) = & \frac{y_2 s_\alpha \pm y_1 c_\alpha}{y_2 s_\beta \pm y_1 c_\beta} \; .
\end{split}
\end{align} 
Note that cancellations can occur, which could make $f_{\pm}$ or $g_{\pm}$ vanish. 
The observed hierarchy in the fermion masses, $m_3 \gg m_2 \gg m_1$, 
may be applied to Eq.~\eqref{eq:newcouplings} to obtain the approximate relations
\begin{align}
\begin{split}
   f_{\pm} \approx  \frac{m_2}{2 m_{2,1}}\left(\frac{c_\alpha}{s_\beta}\mp \frac{s_\alpha}{c_\beta}\right) ,\quad 
   g_{\pm} \approx   \frac{m_2}{2 m_{2,1}}\left(\frac{s_\alpha}{s_\beta}\pm \frac{c_\alpha}{c_\beta}\right) .
\end{split}
\end{align}
For small or large $\tan \beta$ both relations reduce to $f_- \approx \frac{m_2}{m_1} f_+$ and
$g_- \approx \frac{m_2}{m_1} g_+$; meaning that the fermion with a lighter mass ($m_1 < m_2$) has an 
${\cal O}(10-100)$ enhancement in its coupling to the scalars compared to the heavier one.
Moreover, for $\alpha \rightarrow \beta - \pi/2$ all couplings
to the $125 \text{ GeV}$ scalar state, $h$, including the new functions $f_\pm$,
are automatically made SM-like, i.e.\ $\xi_{q_u,q_d,\ell}^h \rightarrow 1$,
while the other couplings end up only depending on $\tan\beta$.  
A further implication of the  alignment limit is that the coupling of the 
$CP$-even state $H$ with the $W$ and $Z$ bosons becomes null.   

\begin{table}
\centering
\begin{tabular}{c|cccc}
  \toprule[0.1em]
   & \multicolumn{4}{c}{Type} \\ 
   & I & II & X & Y   \\
  \hline
  $\xi^h_{q_u}$  & $c_\alpha/s_\beta$ & $c_\alpha/s_\beta$ & $c_\alpha/s_\beta$ & $c_\alpha/s_\beta$  \\
  $\xi^h_{q_d}$ & $c_\alpha/s_\beta$ & $-s_\alpha/c_\beta$ & $c_\alpha/s_\beta$ & $-s_\alpha/c_\beta$  \\ 
  $\xi^h_{\ell}$ & $c_\alpha/s_\beta$ & $-s_\alpha/c_\beta$ & $-s_\alpha/c_\beta$ & $c_\alpha/s_\beta$  \\
  \hline
  $\xi^H_{q_u}$  & $s_\alpha/s_\beta$ & $s_\alpha/s_\beta$ & $s_\alpha/s_\beta$ & $s_\alpha/s_\beta$   \\
  $\xi^H_{q_d}$ & $s_\alpha/s_\beta$ & $c_\alpha/c_\beta$ & $s_\alpha/s_\beta$ & $c_\alpha/c_\beta$   \\ 
  $\xi^H_{\ell}$ & $s_\alpha/s_\beta$ & $c_\alpha/c_\beta$ & $c_\alpha/c_\beta$ & $s_\alpha/s_\beta$  \\
  \hline
  $\xi^A_{q_u}$  & $\cot \beta$ & $\cot \beta$ & $\cot \beta$ & $\cot \beta$   \\
  $\xi^A_{q_d}$  & $-\cot \beta$ & $\tan \beta$ & $-\cot \beta$ & $\tan \beta$   \\ 
  $\xi^A_{\ell}$ & $-\cot \beta$ & $\tan \beta$ & $\tan \beta$ & $-\cot \beta$ \\
  \hline
  $\xi^{H^+}_{q_u}$  & $\cot \beta$ & $\cot \beta$ & $\cot \beta$ & $\cot \beta$   \\
  $\xi^{H^+}_{q_d}$ & $\cot \beta$ & $-\tan \beta$ & $\cot \beta$ & $-\tan \beta$   \\ 
  $\xi^{H^+}_{\ell}$ & $\cot \beta$ & $-\tan \beta$ & $-\tan \beta$ & $\cot \beta$ \\
  \bottomrule[0.1em]
\end{tabular}
\caption{Flavor universal Yukawa couplings of the charged fermions to the Higgs bosons $h,H,A,$ and $H^+$ 
in the conventional 2HDMs with only $\mathbb{Z}_2$. 
\label{tab:couplings1} }
\end{table}

\begin{table}
\centering
\begin{tabular}{c|c}
  \toprule[0.1em]
  & $\mathbb{Z}_3 \rtimes \mathbb{Z}_2$ model   \\
  \hline
  $\xi^h_t$ & $c_\alpha/s_\beta$ \\
  $\xi^h_{b,e}$ & $-s_\alpha/c_\beta$  \\ 
  $\xi^h_{\tau,s,c}$ & $f_+(\alpha,\beta,y_1,y_2)$  \\
  $\xi^h_{\mu,d,u}$ & $f_-(\alpha,\beta,y_1,y_2)$  \\
  \hline
  $\xi^H_t$   & $s_\alpha/s_\beta$  \\
  $\xi^H_{b,e}$ & $c_\alpha/c_\beta$  \\ 
  $\xi^H_{\tau,s,c}$  & $g_+(\alpha,\beta,y_1,y_2)$  \\
  $\xi^H_{\mu,d,u}$  & $g_-(\alpha,\beta,y_1,y_2)$  \\
  \hline
  $\xi^A_t$  & $\cot \beta$  \\
  $\xi^A_{b,e}$  & $\tan \beta$  \\ 
  $\xi^A_{\tau,s,c}$  & $f_+(\beta,\beta,y_1,y_2)$  \\
  $\xi^A_{\mu,d,u}$  &  $f_-(\beta,\beta,y_1,y_2)$  \\
  \hline
  $\xi^{H^+}_t$  & $\cot \beta$  \\
  $\xi^{H^+}_{b,e}$  & $-\tan \beta$  \\ 
  $\xi^{H^+}_{\tau,s}$  & $-f_+(\beta,\beta,y_1,y_2)$  \\
  $\xi^{H^+}_{\mu,d}$  &  $-f_-(\beta,\beta,y_1,y_2)$\\
  $\xi^{H^+}_{c}$  & $f_+(\beta,\beta,y_1,y_2)$ \\
  $\xi^{H^+}_{u}$  &  $f_-(\beta,\beta,y_1,y_2)$  \\
  \bottomrule[0.1em]
\end{tabular}
\caption{Flavor nonuniversal Yukawa couplings, 
cf.\ Eqs.\ (\ref{eq:couneu}, \ref{eq:coucha}), of the charged fermions to the Higgs bosons $h,H,A$, and $H^+$ 
in the 2HDM with $\mathbb{Z}_3 \rtimes \mathbb{Z}_2$. We have grouped the couplings
into different sets with equal or similar characteristics.
\label{tab:couplings2} }
\end{table}

\begin{figure}[t]
\begin{center}
	\includegraphics[scale=0.36]{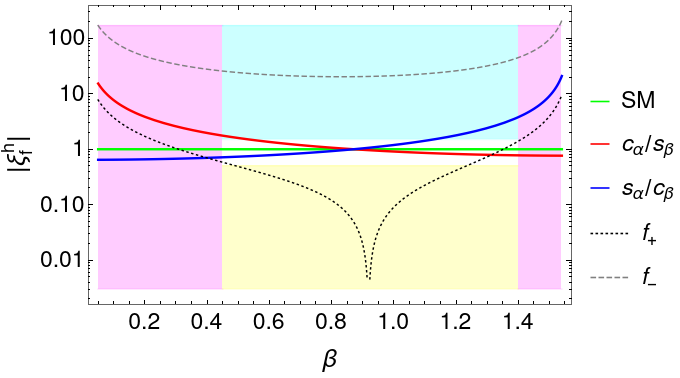}
	\caption{Effective couplings to the $125 \text{ GeV}$ scalar ($h$), as a function of $\beta$, in 
	any of the 2HDM types and the $\mathbb{Z}_3 \rtimes \mathbb{Z}_2$ model. We considered $\alpha=0.7$.
	The blue and red continuous lines represent any of the conventional 2HDMs couplings with NFC whereas the	black (dotted) and gray (dashed) lines the new couplings 
	of our model. The left and right (magenta) shaded regions depict the small and large $\tan \beta$
	limits while the upper (cyan) and lower (yellow) regions 
	the philic ($|\xi^h_f| > 1.5$) and phobic ($|\xi^h_f| < 0.5$) limits,
	respectively. The middle (green) line  the SM limit. 
	Note that the funnel is a consequence of plotting the absolute values of the coupling $\xi^h_f$.
}
	\label{fig:hCoups}
\end{center}
\end{figure}

\begin{figure}[t]
\begin{center}
		\includegraphics[scale=0.36]{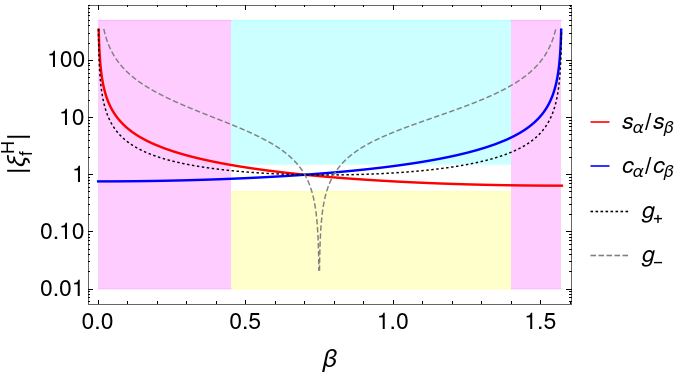}
	\caption{Effective couplings to the heavy $CP$-even scalar ($H$), as a function
	of $\beta$, in any of the 2HDM types and the $\mathbb{Z}_3 \rtimes \mathbb{Z}_2$ model. 
	We considered $\alpha=0.7$.
	The blue and red continuous lines represent any of the conventional
	2HDMs couplings with NFC whereas the black (dotted) and gray (dashed)
	lines the new couplings of our model. The left and right (magenta) 
	shaded regions depict the small and large $\tan \beta$
	limits while the upper (cyan) and lower (yellow) regions the 
	philic ($|\xi^H_f| > 1.5$) and phobic ($|\xi^H_f| < 0.5$) limits,
	respectively. 
	Note that the funnel is a consequence of plotting the absolute values of the coupling $\xi^H_f$.
	}
	\label{fig:HCoups}
\end{center}
\end{figure}

\begin{figure}[t]
\begin{center}
		\includegraphics[scale=0.36]{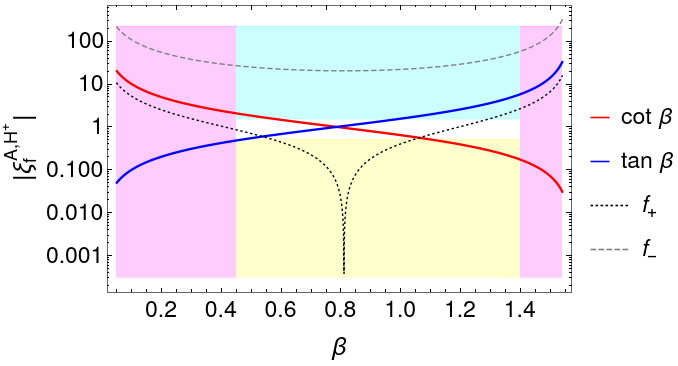}
	\caption{Effective couplings to the neutral $CP$-odd ($A$) 
	and charged ($H^+$) scalars, as a function
	of $\beta$, in any of the 2HDM types and the $\mathbb{Z}_3 \rtimes \mathbb{Z}_2$ model. 
	The blue and red
	continuous lines represent any of the conventional 2HDMs couplings with NFC whereas the
	black (dotted) and gray (dashed) lines the new couplings of our model. The left and right (magenta) 
	shaded regions depict the small and large $\tan \beta$
	limits while the upper (cyan) and lower (yellow) regions 
	the philic ($|\xi^{A,H^+}_f| > 1.5$) and phobic ($|\xi^{A,H^+}_f| < 0.5$)
	limits, respectively.
	Note that the funnel is a consequence of plotting the absolute values of the
	coupling $\xi^{A,H^+}_f$. 
	}
	\label{fig:AHchCoups}
\end{center}
\end{figure}

The resulting couplings have been grouped into different sets corresponding to similar 
characteristics in Table~\ref{tab:couplings2}. 
This also holds for couplings which depend on the Yukawa parameters (and therefore, to the different fermion masses), 
like $\xi^H_{f} = g_+(\alpha,\beta,y_1^f,y_2^f)$ for $f=\tau,s,c$. As they have the same functional dependence 
they are grouped under the category  $\xi^H_{\tau,s,c}$.

In general, conventional 2HDMs with NFC have a moderate behaviour for moderate values of $\tan \beta$.
Their main differences appear in the small (or large) $\tan \beta$ limits. For example, take
the couplings to the charged scalar, $H^+$. In the type-II scenario, its coupling to $t\overline{b}$
is large (philic) at large $\tan \beta$, whereas in the same limit, it is always small (phobic) for the type-I case. In contrast to this typical situation, the $\mathbb{Z}_3 \rtimes \mathbb{Z}_2$ model
shows already at moderate values of $\tan\beta$ either phobic or philic behaviour, see Figs.~\ref{fig:hCoups}-\ref{fig:AHchCoups}. Also it can be seen that for a given value of $\beta$
a given fermion may completely decouple from one of the four scalars and accidentally become inert to
that scalar.

\section{Completing mixing as a guide for new physics}
\label{sec:AddRealScalar}\noindent
While possessing attractive features, the minimal $\mathbb{Z}_3 \rtimes \mathbb{Z}_2$ 
model presented so far does not fully reproduce the fermion mixing and masses. We take
this 'incomplete mixing' as a hint pointing towards new physics.
In the quark sector, the vanishing mixing points to the introduction of a $\text{dim}>4$ operator
that generates small corrections. In a similar fashion, the Majorana
nature of neutrinos could allow dim-4 operators and therefore large contributions to mixing.
The simplest possibility is obtained by introducing a real singlet scalar field, 
which is assumed to transform under $D_3$ as a doublet,
\begin{align}
	\eta \sim {\bf 2} \;.
\end{align}
This field acquires a VEV
\begin{align}
    \langle \eta \rangle = \begin{pmatrix}
        w_{1} \\ 
        w_{2}
    \end{pmatrix}.
\end{align}
Note that by introducing $\eta$ and its non-renormalizable interactions 
we have allowed at tree level the appearance of FCNC. We may assume 
a large mass and later decouple it from the theory. 
While perturbing 2HDMs is typically done to explain anomalies~\cite{Crivellin:2015hha,Crivellin:2017upt}, here 
we need it to complete fermion mixing. Note however that our approach uses an explicit model, i.e.\ 
the symmetry and field content of our model determines the type of Yukawa matrices to be added. 
At last, notice that integrating out the singlet scalar means that our theory 
has become a 2HDM of Type III. We will later demonstrate that the model can be easily made flavor-safe. 
An explicit numerical example will be provided in Sec.\ \ref{ssec:benchmark}. 

\subsection{Quark mixing}
\label{ssec:qmixing}\noindent
In the quark sector, the non-renormalizable dim-5 operators leading to a correct CKM matrix requires a complete UV formulation to be realized. As a simple example that serves as a plausibility argument, consider the following dim-5  effective interactions, invariant under the SM gauge group and the flavor symmetry: 
\begin{align}
\begin{split}
  - \Delta {\cal L}^Q_Y  = & \; 
   \frac{g_1^d}{\Lambda} [\overline{Q}_{L,D} \eta]_{-} d_{R,3} 
   {\Phi}_{2} 
   + \frac{g_2^d}{\Lambda} [\overline{Q}_{L,D} \eta]_{+} d_{R,3}  {\Phi}_{1} 
  \\
  & + \,
    \frac{g_3^d}{\Lambda} \overline{Q}_{L,3} [d_{R,D} \eta]_{+}  {\Phi}_{2}  
    +
    \frac{g_4^d}{\Lambda} \overline{Q}_{L,3} [d_{R,D} \eta]_{-}  {\Phi}_{1} 
    \\
    & + \, \frac{g_5^d}{\Lambda} [\overline{Q}_{L,D} d_{R,D} \eta]_{+}  {\Phi}_{2}
   + \frac{g_6^d}{\Lambda} [\overline{Q}_{L,D} d_{R,D} \eta]_{-}  {\Phi}_{1} \\
    & + \, \text{ H.c.} 
\end{split}
\end{align}
These contributions give rise to small corrections 
in quark mixing through perturbations to the down
quark mass matrix of the form
\begin{align} \label{eq:Pert0}
\Delta {\bf M}_d =
    \begin{pmatrix}
        \epsilon_1 & \epsilon_2 & \epsilon_3 \\
        \epsilon_2 & -\epsilon_1 & \epsilon_4\\
        \epsilon^\prime_3 & \epsilon^\prime_4 & 0
    \end{pmatrix} ,
\end{align}
i.e. the 6 new parameters $g_i^d$, the scale of the dimension-5 operators $\Lambda$ and the two new vevs $\langle \eta_i \rangle = \omega_i$ can be absorbed into 4 parameters $\epsilon_i$ in the down quark mass matrix, which are enough to perturb our initial identity matrix and reproduce the CKM mixing. These effective operators can be realized in an UV-complete model just by adding a vector-like pair of coloured particles with the same gauge quantum numbers as the right-handed down quarks.

In the basis where ${\bf M}_u$ and ${\bf M}_d$ are diagonal,
the perturbation matrix becomes
\begin{align} \label{eq:Pert1}
\begin{split}
    \Delta \widetilde{\bf M}_d = 
    \begin{pmatrix}
    0 & -(\epsilon_1 - i \epsilon_2 ) & 
    \frac{\epsilon_3 - i \epsilon_4}{\sqrt{2}} \\
    \epsilon_1 + i \epsilon_2 & 0 &  \frac{-i(\epsilon_3 + i \epsilon_4)}{\sqrt{2}} \\
    \frac{i(\epsilon_5 + i \epsilon_6)}{\sqrt{2}} &
     \frac{-(\epsilon_5 - i \epsilon_6)}{\sqrt{2}} & 0
    \end{pmatrix} .
\end{split}
\end{align}
Recall that here we still have trivial quark mixing. In order to
obtain a realistic mixing scenario, the perturbations need to be 
sufficiently small compared to the bottom quark mass
but large enough compared to the down and strange quark masses. This implies
that the Yukawa parameters $y_{1,2}^d$ are no longer 
completely satisfying Eq.~\eqref{eq:hierarchy}. Through a qualitative analysis
we find that for
\begin{align}\label{eq:perturbations}
\begin{split} 
    \widetilde{\bf M}_d = \, & \text{diag}( |y_2^d| v_2 - y_1^d v_1 ,
     |y_2^d| v_2 + y_1^d v_1, y_b v_1) 
     + \Delta \widetilde{\bf M}_d \\ \sim &
    \begin{pmatrix}
        m_d & \lambda m_s & \lambda^3 m_b \\
        \lambda m_s & m_s & \lambda^2 m_b \\
        - & - & m_b
    \end{pmatrix} ,
\end{split}
\end{align}
where $\lambda\simeq 0.225$, it is possible to fully reproduce quark mixing without introducing
unacceptably large amounts of flavor violation at tree level. The
$(3,1)$ and $(3,2)$ matrix elements could be taken as zero or of the
same order that their transpose counterparts. 
On the other hand, all entries
are given up to ${\cal O}(1)$ complex factors.
It is interesting
to note that Eq.~\eqref{eq:perturbations} shows an approximate
$U(2)$ flavor symmetry for the first two generations, $m_b \gg m_{d,s}$ (analogously for the up-type quarks).
The above resulting mass matrix is a similar realization of the
'flavorful' 2HDMs investigated in Ref.~\cite{Altmannshofer:2018bch}
wherein Yukawa couplings, for all  charged fermions, 
are chosen as to approximately preserve a $U(2)^5$ 
flavor symmetry acting on the first two generations.

Alternatively, we could have introduced perturbations
through the up-type quarks; however, to reproduce the CKM
mixing would have required a larger modification of the initial
Yukawa parameters, $|y_{2}^u|$ and $y_1^u$, by at least one
order of magnitude. This may be easily appreciated by considering
that a perturbation to the $1-2$ sector of the size $\sqrt{m_1 m_2}$
is enough in the down quark sector, $\sqrt{m_d m_s} \sim 10 \text{ MeV}$, to generate Cabibbo mixing, while
for the up-type quarks it would still require an additional order of magnitude, ${\cal O}(10)\sqrt{m_u m_c}\sim 100 \text{ MeV}$, plus
some extra tuning in the Yukawa parameters to get the correct light quark masses, $m_u$ and $m_c$.

\subsection{Lepton mixing}
\label{ssec:lmixing}\noindent

In the lepton sector, the dominant perturbation contributions 
come through the right-handed neutrinos,
\begin{align}
  - \Delta {\cal L}^{N}_Y  =
 \frac{1}{2} g_1^N \bar{N}^c_{R,1} [\eta N_{R,D}]_{-} + g_2^N [\eta [\bar{N}^c_{R, D} N_{R, D}]_2]_{+} + \text{ H.c.},
\end{align}
producing 
\begin{align} \label{eq:RHN}
 {\Delta} {\cal M}_R = 
 \begin{pmatrix}
	0 & \delta_{N1} & r \delta_{N1} \\
	 \delta_{N1} & r \delta_{N2} & \delta_{N2} 	\\
	r \delta_{N1} &  \delta_{N2} & -r \delta_{N2} 
 \end{pmatrix} ,
\end{align} 
where we have defined $r = \omega_1/\omega_2$, $\delta_{N1} = g_1^N \omega_2$ and $\delta_{N2} = g_2^N \omega_1$ i.e. we can rewrite the 4 new parameters given by $g_1^N$, $g_2^N$, $\omega_1$ and $\omega_2$ in terms of only 3: $r$, $\delta_{N1}$ and $\delta_{N2}$.

The charged lepton contribution remains untouched by the addition of the scalar $\eta$ and is given by Eq.~\eqref{eq:UL}. Once we consider the contributions to the mass matrix of the right-handed neutrinos shown in Eq.~\eqref{eq:RHN} the initial lepton mixing given by Eq.~\eqref{eq:PMNSini} gets modified. If the Yukawa couplings appearing in the neutrino mass matrix are taken real then we have
\begin{equation}
\begin{gathered}
{\bf U}_{L}^e = {\bf U}_{23}(\pi/4, \pm \pi/2) \,,\\
{\bf U}_L^{\nu} = {\bf O}_{23}(\theta_{23}^\nu) {\bf O}_{13}(\theta_{13}^\nu) {\bf O}_{12}(\theta_{12}^\nu)\,,
\end{gathered}
\end{equation}
where ${\bf O}_{ij}(\theta)$ is the usual rotation matrix in the $(i,j)$ plane. It can be shown that
\begin{equation}
    {\bf U}_{23}(\pi/4, \pm \pi/2) {\bf O}_{23}(\theta_{23}^\nu) = {\bf P} \cdot {\bf U}_{23}(\pi/4, \mp \pi/2) \,,
\end{equation}
where ${\bf P}$ is a diagonal unitary matrix which is unphysical. Therefore, if the neutrino sector is real we obtain {\it cobimaximal mixing} \cite{Ma:2015fpa} with $\theta_{23} = \pi/4$ and $\delta_{CP} = \pm \pi/2$ in the lepton sector.  While the sign of $\delta_{CP}$ is not fixed, data seems to favor the negative option \cite{Abe:2019vii}. 
Note that this is a particular case of the general theorem derived in Ref.~\cite{He:2015xha}, i.e.\ if cobimaximal mixing is present in the charged lepton sector and the neutrino sector is real, then the full PMNS matrix is also cobimaximal. In particular, the full lepton mixing parameters are given by
\begin{equation}
\begin{gathered}
\theta_{12} = \theta_{12}^\nu \; ,\quad
\theta_{13} = \theta_{13}^\nu \; ,\quad
\theta_{23} = \pi/4 \; ,\\
\delta_{CP} = \pm \pi/2 \; ,\quad
\phi_{12} =\, 0, \pi/2 \,= \phi_{13}  \;, 
\end{gathered}
\end{equation}
irrespective of $\theta_{23}^\nu$. That is, the large hierarchy between the charged lepton masses coupled with the assumption that the neutrino Yukawas are real leads to cobimaximal mixing i.e.\ maximal atmospheric mixing angle and $\delta_{CP} = \pm \pi/2$. For the other two mixing angles $\theta_{12}$ and $\theta_{13}$ no predictions can be made, but the parameters can be chosen in such a way that they lie inside the experimental constraints. Moreover, the Majorana phases relevant for neutrinoless double beta decay maintain their $CP$ conserving values.

We remark that of course there is no need to assume the neutrino sector to remain real, in the most general scenario  with complex parameters there is enough freedom to fit all the mixing parameters. 
The assumption that the neutrino Yukawas are real, while the charged lepton Yukawas are forced to be complex due to hierarchical masses, may seem ad-hoc but can actually be justified in many different scenarios. For example in Ref.~\cite{Ma:2015pma} the author derives a general loop mechanism in which the neutrino mass matrix is complex but diagonalized by a real orthogonal matrix. This same mechanism could be applied here by changing the type I seesaw neutrino mass generation by an inverted loop seesaw mediated by three real scalars. Then, the cobimaximal nature of the PMNS would remain. 
Another option would be to explicitly impose a remnant $CP$ symmetry in the neutrino sector.\\

It is worth to note that our scenario is minimal and quite simple, yet, it leads to such a restricted scenario. The SM symmetry group is extended by just $D_3$ while the particle content is enlarged by an extra Higgs gauge doublet and an $D_3$ doublet $\eta$ which is a gauge singlet.

\subsection{The scalar potential}
\label{ssec:scalarpot}
\noindent
The most general scalar potential invariant under $\mathbb{Z}_3 \rtimes \mathbb{Z}_2$ is 
\begin{equation}
    V = V_\text{2HDM}^{\mathbb{Z}_2} + V_{\eta}^{\mathbb{Z}_3 \rtimes \mathbb{Z}_2} + V_{\Phi \eta}^{\mathbb{Z}_3 \rtimes \mathbb{Z}_2} \;,
\end{equation}
with the first term given in Eq.~\eqref{eq:2HDMZ2} and 
\begin{align} \label{eq:goldstone}
\begin{split}
V_\eta^{\mathbb{Z}_3 \rtimes \mathbb{Z}_2} & =   
\frac{\mu^2_\eta}{2} [\eta \eta]_+
+\frac{\lambda_{\eta 1}}{2} [\eta \eta]_+^2
+\frac{\lambda_{\eta 2}}{2} \Big[ [\eta \eta]_2 [\eta \eta]_2 \Big]_+ \;,
\\
V_{\Phi \eta}^{\mathbb{Z}_3 \rtimes \mathbb{Z}_2} & =   \left[ \zeta_{1}(\Phi^\dagger_1 \Phi_1)
+ \zeta_{2} (\Phi^\dagger_2 \Phi_2) \right] (\eta \eta)_+ \;,
\end{split}
\end{align}
where $[\eta \eta]_k$ represents one of the three possible choices (${\bf 1}_+$, ${\bf 1}_-$, ${\bf 2}$), 
all couplings are real (due to hermiticity) ensuring a $CP$ conserving potential 
and we have omitted those terms involving $[\eta \eta]_-$ as it is zero.
This potential has an extra Goldstone boson due
to the fact that $[\eta \eta]_+ = \eta_1^2 + \eta_2^2$ is equivalent to
$(\eta_1 - i\eta_2)(\eta_1 + i\eta_2)$ and $[[\eta \eta]_2 [\eta \eta]_2]_+ = ([\eta \eta]_+)^2$. 
Then, Eq.~\eqref{eq:goldstone} is accidentally invariant
under a global $U(1)$ symmetry originated
from the two components of the flavor doublet scalar, $\eta$.
To avoid its appearance we softly-break the accidental symmetry by introducing
\begin{align}
    V_\text{soft} = \frac{\mu^2_{1}}{2} \eta_1^2 
    +\frac{\mu^2_{2}}{2} \eta_2^2 -\mu^2_{12} \eta_1 \eta_2 \;.
\end{align}
Additionally, the fact that the heavy quark masses
are simply given by $m_t \simeq y_t v_2$ and
$m_b \simeq y_b v_1$ naturally points to having order one Yukawas and hierarchical VEVs in the
range
\begin{align}
    v_2 \simeq v \qquad \text{and} \qquad v_1 \sim 
    [ 1 ,10 ] \, \text{ GeV} \; , 
\end{align}
meaning that $\tan \beta \in (10,100)$.
To create such a hierarchy while maintaining all scalar masses around
the electroweak scale we need to consider $m_{22}^2 <0$, $m_{11}^2 >0$, and introduce the soft-breaking term 
\begin{align}
    - m_{12}^2 (\Phi_1^\dagger \Phi_2 + \text{ H.c.}) \;,
\end{align}
where $m_{12} \sim {\cal O}(10) \text{ GeV}$.
By assuming $|m_{11}|, |m_{22}| \sim 100 \text{ GeV}$, a straightforward calculation then leads to
\begin{align}
    v_2 \simeq \sqrt{\frac{-m_{22}^2}{\lambda_2}} \quad
    \text{ and }
    \quad
    v_1 \simeq \frac{m_{12}^2 v_2}{m_{11}^2 + \lambda_{345} v_{2}^2} \;.
\end{align}
The smallness of $v_1$ is thus natural as one recovers a larger symmetry when setting it to zero. 

The minimization conditions read
\begin{align}
\begin{split} \nonumber
    - m_{11}^2 & = \lambda_1 v_1^2 + \lambda_{345} v_2^2 +  (w_1^2 + w_2^2) \zeta_2 - m_{12}^2 \tan \beta \;, \\ \nonumber
    - m_{22}^2 & = \lambda_2 v_2^2 + \lambda_{345} v_1^2 +  2 (w_1^2 + w_2^2) \zeta_1 
    - m_{12}^2 \cot \beta
    \;, \\ \nonumber
    - \mu_\eta^2 & = 2 (w_1^2 + w_2^2) \bar{\lambda} + 2v_1^2 \zeta_2 + 2v_2^2 \zeta_1 +\mu_2^2-\frac{w_1}{w_2}\mu_{12}^2 \;, \\ \nonumber
    - \mu_\eta^2 & = 2 (w_1^2 + w_2^2) \bar{\lambda} + 2v_1^2 \zeta_2 + 2v_2^2 \zeta_1 +\mu_1^2-\frac{w_2}{w_1}\mu_{12}^2 \;, \nonumber
\end{split}
\end{align}
where $\bar{\lambda} = \lambda_{\eta 1} + \lambda_{\eta 2}$. The latter two conditions can only be met if
\begin{equation}
    \mu_{12}^2 = -\frac{w_1 w_2}{w_1^2 - w_2^2}(\mu_1^2-\mu_2^2) \;.
\end{equation}
The general expressions for the squared mass matrices are given in Appendix~\ref{app:ScalarPot}.

In order to decouple $\eta$ from the 2HDM 
we assume its mass (or VEV) to be large enough and $\zeta_{1,2} \rightarrow 0$. Then,
for the full potential, $V + V_\text{soft}$, to be bounded from below we require the well-known relations
\begin{equation}
\begin{gathered}
    \lambda_1 \geq 0 \;, \quad  \lambda_2 \geq 0 \;, \quad
    \lambda_3 \geq - \sqrt{\lambda_1 \lambda_2} \;, \\
    \lambda_3 + \lambda_4 - |\lambda_5| \geq - \sqrt{\lambda_1 \lambda_2}\;,
\end{gathered}
\end{equation}
while for the new contributions
\begin{align}
\begin{split}
\lambda_{\eta 1} \geq 0   \quad \text{and} \quad \lambda_{\eta 2} \geq 0 \;,
\end{split}
\end{align}
which all  are sufficient and necessary conditions.

\subsection{Numerical example}
\label{ssec:benchmark}
\noindent

In the following, we give a numerical example of how the perturbations brought by 
the addition of $\eta$ modify our initial 2HDM setup. We assign a best-fit value to our set of complex parameters
$\{\epsilon_1, \epsilon_2,\epsilon_3,\epsilon_4,\epsilon_5,\epsilon_6\}$ by virtue of a $\chi^2$ 
fit to the three down quark masses and four quark mixing parameters
\begin{equation}
\begin{split}
\chi ^{2} = &\sum_{f=d,s,b}\frac{(m_{f}^{\text{th}}-m_{f}^{\text{exp}})^{2}}{\sigma_{f}^{2}}
+\frac{(|\mathbf{V}_{12}^{\text{th}}|-|\mathbf{V}_{12}^{\text{ckm}}|)^{2}}{\sigma _{12}^{2}}
\\
& +\frac{(|\mathbf{V}_{23}^{\text{th}}|-|\mathbf{V}_{23}^{\text{ckm}}|)^{2}}{\sigma _{23}^{2}}+\frac{(|\mathbf{V}_{13}^{\text{th}}|-|\mathbf{V}_{13}^{\text{ckm}}|)^{2}}{\sigma _{13}^{2}}
\\
& + \frac{(J_{q}^{\text{th}}-J_{q}^{\text{exp}})^{2}}{\sigma _{J}^{2}}\, \;,
\end{split}
\end{equation}
where the value of the masses is taken at the $Z$ boson mass scale, $M_Z$, 
using the \texttt{RunDec} package~\cite{Herren:2017osy}
\begin{align}
\begin{split}
    m_d^{\text{exp}} (M_Z) & = 0.0027^{+0.0003}_{-0.0002} \text{ GeV} \;, \\
    m_s^{\text{exp}} (M_Z) & = 0.055^{+0.004}_{-0.002} \text{ GeV} \;, \\
    m_b^{\text{exp}} (M_Z) & = 2.86 \pm 0.02 \text{ GeV}  \;,
\end{split} 
\end{align}
and
\begin{align}
\begin{split}
    |\mathbf{V}_{12}^{\text{ckm}}| & = 0.22452 \pm 0.00044 \;, \\
    |\mathbf{V}_{23}^{\text{ckm}}| & = 0.04214 \pm 0.00076  \;, \\
    |\mathbf{V}_{13}^{\text{ckm}}| & = 0.00365 \pm 0.00012 \;,\\
    J_{q}^{\text{exp}} & = (3.18\pm 0.15) \times 10^{-5}\;,
\end{split}
\end{align}
as shown in the most recent global fit from the PDG~\cite{Tanabashi:2018oca}.
As a proof of principle, we consider a minimal scenario with the least 
number of parameters. We assume all of them real except for
$\epsilon_4$ which we consider it as purely imaginary and set $\epsilon_{5,6} = 0$. Also we allow for small variations in the initial down quark Yukawa 
couplings appearing in Eq.~\eqref{eq:masses}. 

The following best-fit values, 
\begin{equation}
\begin{gathered}
    \epsilon_1 = 4.4634 \text{ MeV} \;, \qquad
    \epsilon_2 = 12.1428 \text{ MeV} \;, \\
    \epsilon_3 = 103.6520 \text{ MeV} \;, \qquad
    \epsilon_4 = -i \, 60.9786 \text{ MeV} \;,
\end{gathered}
\end{equation}
reproduce the down quark masses and the observed CKM mixing at the 
$1 \sigma$ level with a quality of fit of $\chi^2_\text{d.o.f.} = 0.49$.

Besides their role in mixing, the introduction of perturbations has also brought
FCNC at tree level. We now show how the size of the contributions is still 
sufficiently small. Through the best-fit values we calculate the unitary
transformations for the left- and right-handed fields. With them the corresponding
down quark Yukawa matrices, in the mass basis, are
\begin{align} \label{eq:YukFCNC}
\begin{split}
    \widetilde{\bf \Gamma}_{1} & \lesssim 
    \begin{pmatrix}
        10^{-4} & 10^{-4} & 10^{-7}\\
        10^{-4} & 10^{-3} & 10^{-5}\\
        10^{-5} & 10^{-5} & 10^{-1}
    \end{pmatrix} + i
    \begin{pmatrix}
        10^{-9} & 10^{-7} & 10^{-7}\\
        10^{-7} & 10^{-8} & 10^{-6}\\
        10^{-6} & 10^{-5} & 10^{-9}
    \end{pmatrix},
    \\
    \widetilde{\bf \Gamma}_{2} & \lesssim
    \begin{pmatrix}
        10^{-4} & 10^{-5} & 10^{-8}\\
        10^{-5} & 10^{-4} & 10^{-7}\\
        10^{-6} & 10^{-6} & 10^{-9}
    \end{pmatrix} 
    + i
    \begin{pmatrix}
        10^{-10} & 10^{-8} & 10^{-8}\\
        10^{-8} & 10^{-8} & 10^{-8}\\
        10^{-7} & 10^{-6} & 10^{-10}
    \end{pmatrix},
\end{split}
\end{align}
where we have assumed $v_1 \sim 10 \, m_b$ and $v_2 \sim m_t$ to estimate 
the upper bounds and which are all consistent with those presented
in Refs.~\cite{Crivellin:2013wna,Altmannshofer:2018bch}. 
There are in fact three different scenarios from which
Eq.~\eqref{eq:YukFCNC} represents one of them. As all the independent perturbations defined
in Eq.~\eqref{eq:Pert0} originate from both Higgs doublets, $\Phi_1$ and $\Phi_2$,
we can define three different benchmark scenarios as follows: all the perturbations come from
i) $\Phi_1$, ii) $\Phi_2$, or iii) both. Our choice in Eq.~\eqref{eq:YukFCNC} depicts
the first case. We left for future work a detailed study of the differences between this approach and
the conventional 2HDMs.

\section{Conclusions}
\label{sec:concl}
\noindent
We have considered a new class of 2HDM where the conventional $\mathbb{Z}_2$ 
symmetry, by which FCNC can be naturally avoided, has been enlarged to
${\cal F}\rtimes \mathbb{Z}_2$ such that symmetry constrains the Yukawa sector 
but goes unnoticed by the scalar sector. 
In particular, we have shown that the minimal case with $ \mathbb{Z}_3 \rtimes \mathbb{Z}_2$  
is able to provide trivial quark and maximal atmospheric mixing at leading order. 
A further implication to this class of models is that couplings between the
fermions to the scalars are nonuniversal, compared to the conventional types
where couplings are universal. At last, we have taken the incompleteness of 
fermion mixing as a hint pointing towards new physics. To this end we have 
included two real scalar gauge singlets which transform as a flavor doublet, 
and are later integrated out by assuming them to be properly heavy. We have shown that 
quark mixing can be set in agreement with the latest global fits while the 
lepton mixing can become cobimaximal, i.e.\ maximal atmospheric mixing 
and maximal $CP$ violation. 
We have treated the introduction of the real scalars as
a new way of adding perturbations to 2HDMs in a systematic manner
by demanding them to be invariant under the flavor symmetry.
In general, these additions have the effect of breaking flavor 
conservation and tree level FCNC, mediated by the neutral scalars,
are induced. However, the size of the contributions remains sufficiently
small thanks to the approximate presence of a $U(2)^3$ global
flavor symmetry in the light quark sector.

\section*{ACKNOWLEDGMENTS}
\noindent
The authors thank Andreas Trautner and Rahul Srivastava for useful conversations.
S.C.C's work is supported by FPA2017-85216-P (AEI/FEDER, UE), SEV-2014-0398, PROMETEO/2018/165 (Generalitat  Valenciana), Spanish Red Consolider MultiDark FPA2017-90566-REDC and the FPI grant BES-2016-076643. 
The work of W.R.\ is supported by the DFG with grant RO 2516/7-1 in the Heisenberg program. U.J.S.S.\ acknowledges support from CONACYT (M\'exico). S.C.C would like to thank the Max-Planck-Institute for Nuclear Physics in Heidelberg for their hospitality during his visit, where this work was initiated.

\appendix

\section{Product rules of $D_3 \simeq \mathbb{Z}_3 \rtimes \mathbb{Z}_2$}
\label{app:Rules}\noindent
$D_3$ is the smallest non-Abelian discrete symmetry group. It describes the symmetrical properties of an equilateral triangle. It has three irreducible representations: two singlets, ${\bf 1}_+$ and ${\bf 1}_-$, and one doublet, ${\bf 2}$. Their product rules are
\begin{align}
\begin{split}
	{\bf 1}_+ \times & {\bf 1}_+ = {\bf 1}_+ \;, \qquad {\bf 1}_- \times {\bf 1}_- = {\bf 1}_+ \;, \qquad
	{\bf 1}_- \times {\bf 1}_+ = {\bf 1}_- \;,\\
	{\bf 1}_+ \times & {\bf 2} = {\bf 2} \;, \qquad {\bf 1}_- \times {\bf 2} = {\bf 2} \;, \qquad
	{\bf 2} \times {\bf 2} = {\bf 1}_+ +{\bf 1}_- + {\bf 2} \;.
\end{split}
\end{align} 
In particular, the tensor product of two doublets, ${\bf a} =(a_1 , a_2)^T$ 
and ${\bf b} = (b_1 , b_2)^T$, is explicitly given as
\begin{align}
\begin{split}
	{\bf a} \times {\bf b} = &\; (a_1 b_1 + a_2 b_2 )_{{\bf 1}_+ }
	+
	(a_1 b_2 - a_2 b_1 )_{{\bf 1}_-}
	\\
	& + \,
	\begin{pmatrix}
	 a_1 b_2 + a_2 b_1 \\
	 a_1 b_1 - a_2 b_2 
	\end{pmatrix}_{\bf 2} \;.
\end{split}
\end{align}

\section{Renormalization group equations}
\label{app:RGE}\noindent
In a model with $N$ Higgs doublets, wherein all Higgses 
couple to all fermions, the fermion mass matrices are
expressed as a linear combination of $N$ Yukawa matrices times a VEV.
Given an initial setup, the one-loop renormalization group equations (RGE) 
tell us how stable are the initial mass matrices at higher scales and if new 
flavor structures may appear, giving rise to misalignments. 
The one-loop RGE has been calculated in Ref.\ \cite{Ferreira:2010xe} and reads 
\begin{align}
\begin{split}
    16 \pi^2 \mu \frac{d}{d\mu} \Gamma_k  = & 
 a_\Gamma \Gamma_k \\
   &  + \sum_{l=1}^N \left[ 3\, \text{Tr}(\Gamma_k \Gamma_l^\dagger+\Delta_k^\dagger \Delta_l) +\text{Tr}(\Pi_k \Pi_l^\dagger)\right] \Gamma_l \\
 & + \sum_{l=1}^N \left(-2\Delta_l \Delta_k^\dagger \Gamma_l
    + \Gamma_k \Gamma_l^\dagger \Gamma_l \right) \\
& + \frac{1}{2} \sum_{l=1}^N \left(   \Delta_l \Delta_l^\dagger \Gamma_k +\Gamma_l \Gamma_l^\dagger \Gamma_k\right),
\end{split} 
\end{align}
\begin{align}
\begin{split}
    16 \pi^2 \mu \frac{d}{d\mu} \Delta_k  = & 
 a_\Delta \Delta_k \\
   &  + \sum_{l=1}^N \left[ 3\, \text{Tr}(\Delta_k \Delta_l^\dagger+\Gamma_k^\dagger \Gamma_l) +\text{Tr}(\Pi_k \Pi_l^\dagger)\right] \Delta_l \\
 & + \sum_{l=1}^N \left(-2\Gamma_l \Gamma_k^\dagger \Delta_l
    + \Delta_k \Delta_l^\dagger \Delta_l \right) \\
& + \frac{1}{2} \sum_{l=1}^N \left(   \Gamma_l \Gamma_l^\dagger \Delta_k +\Delta_l \Delta_l^\dagger \Delta_k\right),
\end{split}
\end{align}
\begin{align}
\begin{split}
    16 \pi^2 \mu \frac{d}{d\mu} \Pi_k  = & 
 a_\Pi \Pi_k \\
   &  + \sum_{l=1}^N \left[ 3 \, \text{Tr}(\Delta_k \Delta_l^\dagger+\Gamma_k^\dagger \Gamma_l) +\text{Tr}(\Pi_k \Pi_l^\dagger)\right] \Pi_l \\
 & + \sum_{l=1}^N \left(\Pi_k \Pi_l^\dagger \Pi_l
    + \frac{1}{2} \Pi_l \Pi_l^\dagger \Pi_k \right) ,
\end{split}
\end{align}
where we have followed the notation introduced in~\cite{Ferreira:2010xe}.
Here $\mu$ denotes the renormalization scale and
\begin{align}
\begin{split}
    a_\Gamma =& -8 g_s^2 -\frac{9}{4} g^2 - \frac{5}{12} g^\prime{}^2 \;, \\
    a_\Delta =& -8 g_s^2 -\frac{9}{4} g^2 - \frac{17}{12} g^\prime{}^2 \;, \\
    a_\Pi =& -\frac{9}{4} g^2 - \frac{15}{4} g^\prime{}^2 \;,
\end{split}
\end{align}
where $g_s$, $g$, and $g^\prime$ are the gauge couplings of the SM gauge group,
$SU(3)_C \times SU(2)_L \times U(1)_Y$, respectively.

\section{Squared mass matrices}
\label{app:ScalarPot}
\noindent
The squared mass matrix for the $CP$-even scalars reads
\begin{align}
{\bf M}_{CP-\text{even}}^2 =
\begin{pmatrix}
{\bf m}^2_{\phi \phi} &  {\bf m}^2_{\phi \eta} \\
{\bf m}^2_{\phi \eta} & {\bf m}^2_{\eta \eta} 
\end{pmatrix},
\end{align}
where
\begin{align}
{\bf m}^2_{\phi \phi} & =
\begin{pmatrix}
    2v_2^2 \lambda_2 + m_{12}^2\cot \beta  & 2v_1 v_2 \lambda_{345} -m_{12}^2 \\
 2 v_1 v_2 \lambda_{345} -m_{12}^2 & 2v_1^2 \lambda_1 + m_{12}^2 \tan \beta 
\end{pmatrix},
\end{align}
\begin{align}
 {\bf m}^2_{\phi \eta} & =
\begin{pmatrix}
   2\sqrt{2} \zeta_1 v_2 w_1 & 2\sqrt{2} \zeta_2 v_1 w_1 \\
    2\sqrt{2} \zeta_1 v_2 w_2 & 2\sqrt{2}\zeta_2 v_1 w_2 
\end{pmatrix} ,
\end{align}
\begin{align}
{\bf m}^2_{\eta \eta} & =
\begin{pmatrix}
4\bar{\lambda} w_1 w_2 - \frac{w_1 w_2(\mu_1^2 -\mu_2^2)}{w_2^2 -w_1^2}
    & 4\bar{\lambda} w_2^2   + \frac{w_1^2(\mu_1^2-\mu_2^2)}{w_2^2-w_1^2} \\
  4\bar{\lambda} w_1^2 +\frac{w_2^2(\mu_1^2-\mu_2^2)}{w_2^2-w_1^2} 
   & 4\bar{\lambda} w_1 w_2 - \frac{w_1 w_2(\mu_1^2 -\mu_2^2)}{w_2^2 -w_1^2}
\end{pmatrix} ,
\end{align}
 whereas for the $CP$-odd and charged scalars we have
 \begin{align}
    {\bf M}_{CP-\text{odd}}^2 =
    \begin{pmatrix}
     2v_2^2 \lambda_5 - m_{12}^2 \tan \beta & -2v_1 v_2 \lambda_5 +m_{12}^2 \\
     -2v_1 v_2 \lambda_5 +m_{12}^2 & 2v_1^2 \lambda_5 - m_{12}^2 \cot \beta
     \end{pmatrix} ,
 \end{align}	
and 
 \begin{align}
    {\bf M}_\pm^2 =
    \begin{pmatrix}
     2v_2^2 (\lambda_4 + \lambda_5) - m_{12}^2 \tan \beta & -2v_1 v_2 (\lambda_4 + \lambda_5) +m_{12}^2 \\
     -2v_1 v_2 (\lambda_4 + \lambda_5)  +m_{12}^2 & 2v_1^2 (\lambda_4 + \lambda_5) - m_{12}^2 \cot \beta
     \end{pmatrix} .
 \end{align}
Note that the mass matrices for the $CP$-odd and charged scalars are rank one while rank four 
for the $CP$-even case. 
		
\bibliographystyle{apsrev4-1}		
\bibliography{2HDMFlavZ2.bib}

\end{document}